# Effect of Annealing on Flexoelectricity in Hafnium Oxide (HfO2)

*Daniel Moreno-Garcia[1], Luis Guillermo Villanueva[1]*

[1] Advanced NEMS Group, École Polytechnique Fédérale de Lausanne (EPFL), Lausanne 1015, Switzerland
E-mail: Guillermo.Villanueva@epfl.ch

**Abstract**

Flexoelectricity is universal in all dielectrics, effective at high temperatures, and a promising transduction technique for nanoelectromechanical systems (NEMS). However, as flexoelectricity is still in its early stages, many aspects require further investigation. Understanding how flexoelectricity depends on material parameters like crystallographic phase and how temperature might affect it, is important for selecting and optimizing the right material for technological applications. This work studies the influence of high-temperature annealing (and the consequent crystallization) in the flexoelectricity of hafnium oxide (HfO2), a material with significant technological relevance. We measure the flexoelectric coefficient for amorphous (not annealed) and annealed (slightly crystalline) phases of HfO2, with samples annealed in nitrogen or oxygen atmospheres. Our results indicate that the amorphous phase of HfO2 exhibits the highest flexoelectric coefficient (105 ± 10 pC/m), while annealed samples show a significant decrease, with the lowest value in nitrogen-annealed samples (26 ± 4 pC/m). Samples annealed in an oxygen atmosphere improve flexoelectric properties (54 ± 6 pC/m) compared to those annealed in nitrogen. Using cross-sectional imaging, X-ray diffraction, resonance frequency characterization, and relative permittivity measurements, we find that annealing promotes crystallization into the tetragonal phase and increases internal stress within the HfO2 layer, while most other parameters remain constant. We attribute the differences in flexoelectricity from the annealed samples to the quantity of oxygen vacancies in hafnium oxide. These oxygen vacancies in hafnium oxide seem to negatively affect the flexoelectric coefficient. This finding can be applied to optimize materials to enhance their flexoelectric properties.

Flexoelectricity, the coupling between strain gradient (bending) and polarization, was first theorized by Kogan[1] in 1964 and predicted to be proportional to the material's relative permittivity. Initially, research in this field focused predominantly on liquid crystals[2]. However, in the early 2000s, Ma and Cross[3] measured the flexoelectric coefficient in high-dielectric constant (high-κ) materials, demonstrating that effective transducers could be fabricated using non-piezoelectric materials. This breakthrough has since shifted the spotlight to flexoelectricity as a promising technique for actuation and sensing, particularly as devices scale down to nanometric dimensions[4]. Its potential to outperform piezoelectricity in nanoscale applications[5] positions flexoelectricity as a strong candidate for the advancement of nanoelectromechanical systems (NEMS)[6,7].

Flexoelectricity is inherent to all dielectrics, regardless of their crystallographic symmetry. This universality eliminates the dependency on non-centrosymmetric materials required for piezoelectricity[8], thereby expanding the range of materials that can be used. Additionally, flexoelectric materials bypass the curie temperature limitation that constrains piezoelectric materials. The use of lead-based compounds in some piezoelectric materials poses toxicity issues, whereas flexoelectric devices can be manufactured from simple, non-toxic dielectrics, offering a safer alternative.

To establish flexoelectric materials as a viable technology, it is important to fabricate materials in the nanoscale with a robust flexoelectric response[9]. Achieving this can be approached either by engineering the materials to increase the strain gradient for a given load or by enhancing the material's flexoelectric coefficient.

Increasing the strain gradient for a given load can be accomplished through various strategies. One approach involves engineering materials with nanometrically triangular voids[10], which induce significant additional strain when the material is deformed. Another method combines nanowires with nanosheets to create composites that exhibit strong local curvature[11]. With the same philosophy, significant flexoelectric charges can be obtained with a twisted foam composite made from polydimethylsiloxane (PDMS)[12]. When deformed, this foam composite is capable of generating an electric output at the microampere scale, sufficient to charge small electronic devices.

Enhancing the flexoelectric coefficient itself is another effective strategy. One approach involves using materials with higher relative permittivity, as flexoelectricity appears to depend quadratically on this property[13]. The relative permittivity of ferroelectric materials varies with temperature and reaches its peak near the critical ferroelectric transition. The largest flexoelectric coefficients have been reported close to this phase transition[14–18]. Other studies have shown that doping the material with extrinsic elements can significantly enhance flexoelectricity, sometimes doubling the coefficient[19]. While most flexoelectric measurements have been performed on monocrystalline materials, few studies have reported measurements on non-fully crystalline materials[13,20]. Indeed, the impact of crystallinity on flexoelectricity remains unexplored, which constitutes a significant knowledge gap in optimizing the material choice. If amorphous materials performed comparably or better than their crystalline counterparts, this would significantly expand the range of materials suitable for flexoelectric applications, including glassy materials.

This work aims to provide insights into the difference in flexoelectric coefficients depending on the crystallinity of the material and annealing steps. To ensure precise comparisons, we start with a common amorphous material (hafnium oxide) and anneal selected samples to promote crystallization in the dielectric layer. We then compare the flexoelectric coefficients in all types of samples by fabricating micromechanical cantilevers (clamped-free) and clamped-clamped beams. The devices (cantilevers and C-C beams) are fabricated with a range of lengths (from 6 μm to 22 μm and from 15 μm to 80 μm, respectively) and are made of the dielectric under study ($HfO_2$) sandwiched between metal electrodes for actuation. Figure 1a shows the schematic of the microcantilevers, with 20 nm thick metal (titanium and platinum) electrodes and a 50 nm thick hafnium oxide layer. The hafnium oxide deposition is performed using Atomic Layer Deposition (ALD) at 200 ºC, while the platinum electrodes are deposited via evaporation with a thin (2 nm) titanium layer for adhesion. Figures 1b and 1c display SEM images of the final devices, which have a width of 2.5 μm. The cantilever has a length of 20 μm (Figure 1b), and the clamped-clamped beam is 60 μm long (Figure 1d).

The layer of $HfO_2$ is originally amorphous, as shown in the XRD measurements of Fig. 2b. To obtain devices with different crystallinity, some chips are annealed[21] before defining the beams. The selected annealing temperature profile is depicted in Figure 1c. The temperature is increased at a rate of 20 ºC/min until reaching 700 ºC, where it is held for 15 minutes. Subsequently, the heating is stopped, and the temperature is reduced by convection.

Following the annealing, the fabrication process continues with the patterning and release of the cantilevers. This method enables the fabrication of cantilevers with $HfO_2$ in three distinct forms: amorphous, annealed in a nitrogen ($N_2$) atmosphere, and annealed in an oxygen ($O_2$) atmosphere. Detailed information on the fabrication process flow and the various annealing tests is provided in the supplementary material.

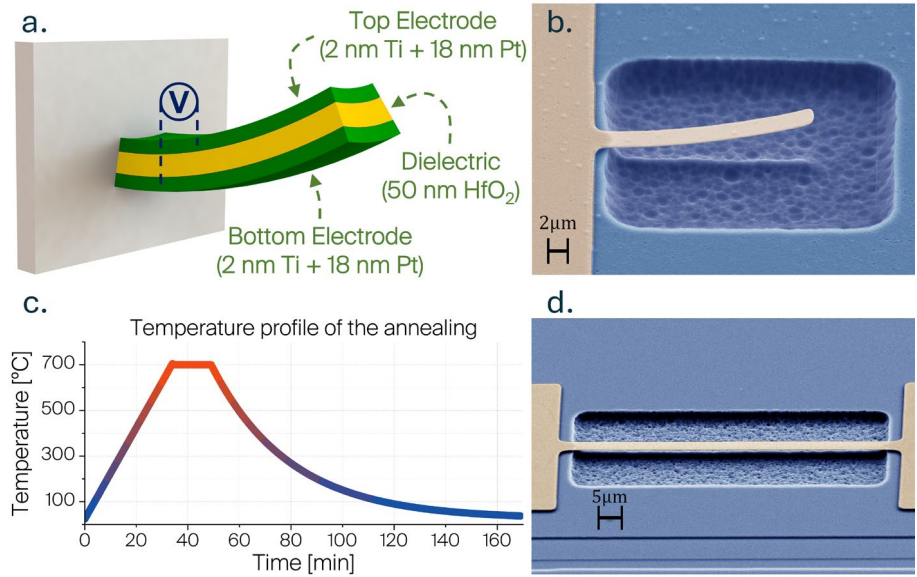

**Figure 1.** a) Schematic of the cantilevers used in flexoelectric measurements, composed of a 2 nm adhesion layer of titanium (Ti) and 18 nm platinum (Pt) sandwiching a 50 nm dielectric $HfO_2$. Both amorphous and annealed devices were fabricated using the same process. b) Scanning Electron Microscope (SEM) image of a fabricated cantilever with annealed $HfO_2$. c) Temperature profile utilized for the annealing of $HfO_2$ samples before cantilever fabrication. For more comprehensive fabrication details, refer to the supplementary material. d) Scanning Electron Microscope image of a fabricated clamped-clamped beam with amorphous $HfO_2$.

Before measuring the flexoelectric coefficient, we analyze the changes that the annealing causes in the material properties such as density, crystallographic phase, Young's modulus and internal stress. These variables are necessary for the accurate measurement and comparison of the flexoelectric coefficients.

Cross-sectional SEM images are taken to verify if the annealing process affects the thickness of the layers (Figure 2a). The measurements show that, regardless of whether the samples are amorphous or annealed, the electrodes and the hafnium oxide have the same thickness. The measurement indicates that no apparent change in the material density occurs for the annealed samples.

Additionally, to confirm that the annealing process alters the crystallographic phase of $HfO_2$, X-ray diffraction (XRD) analysis is performed. Figure 2b illustrates the emergence of hafnium oxide peaks in the annealed samples compared to the amorphous sample. These peaks are located at 34.9º and 35.6º and are associated with the tetragonal phase of $HfO_2$ ($P_{4_2}/nmc$)[22].

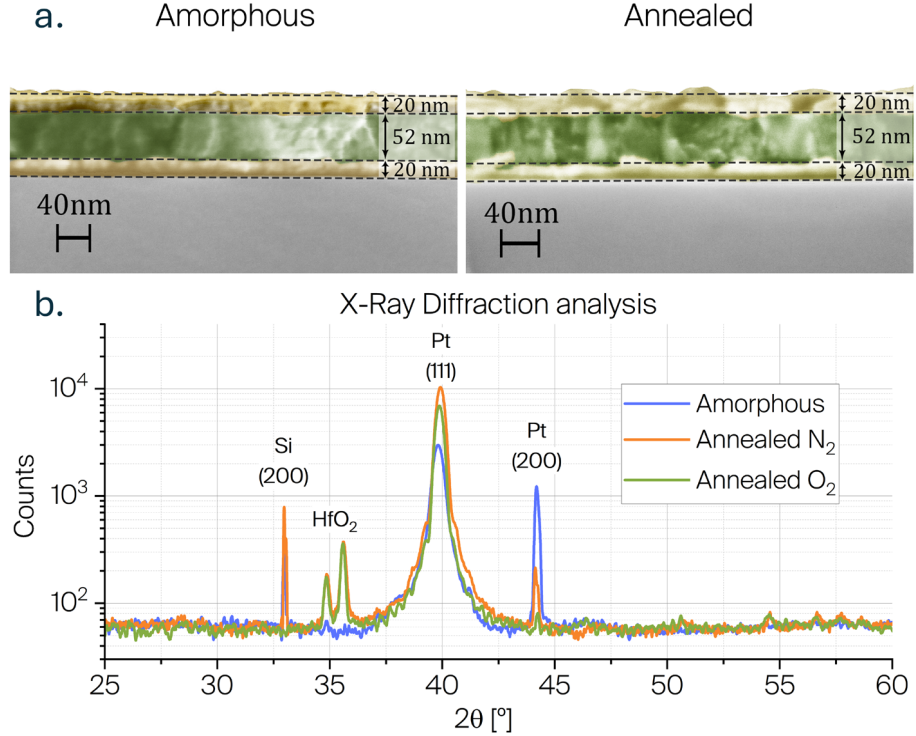

**Figure 2.** a) Cross-sectional images of the amorphous and annealed devices, demonstrating that layer thicknesses remain unchanged during the annealing process. Each device has a 20 nm top and bottom metal electrode sandwiching a 50 nm HfO₂ dielectric layer. b) X-ray diffraction analysis of amorphous and annealed samples. The annealed profiles exhibit the emergence of two new peaks, indicative of the tetragonal phase of HfO₂ crystallization. Peaks corresponding to Silicon (Si) and Platinum (Pt) remain consistent between the amorphous and annealed samples.

To understand the additional effects of annealing on the samples, we monitor the changes in Young's modulus and internal stress by measuring the resonance frequency of the devices. For cantilevers, the resonance frequency inversely depends on the square of the cantilever length[23] ($f_{R,C} \propto t/L^2 \sqrt{E/\rho}$). Other variables are the thickness ($t$), the Young's modulus ($E$) and the cantilever's density ($\rho$). For highly stressed clamped-clamped beams, the resonance frequency is inversely linear with the length of the beams[24] ($f_{R,CC} \propto 1/L \sqrt{\sigma/\rho}$) and depends on the internal stress ($\sigma$).

For cantilevers, the resonance frequencies for both amorphous and annealed samples are similar and align with our finite element method simulations (see Figure 3a). As the resonance does not change, and the geometry remains the same (Figure 2a-b), following the above expression for $f_{R,C}$, we conclude that Young's modulus and density remain constant for both materials' phases.

For clamped-clamped beams, the resonance frequencies observe an increase in the annealed samples. Using the expression for $f_{R,CC}$, we can conclude that the internal stress increases with the annealing and has the same magnitude regardless of the annealing environment (N₂ or O₂). The increase in internal stress can be quantified using finite element method simulations or its theoretical definition ($f_{R,CC}$), to be 700 MPa for the amorphous samples and 1100 MPa for the annealed ones. This represents a net increase of 400 MPa (or 60%) for the annealed samples.

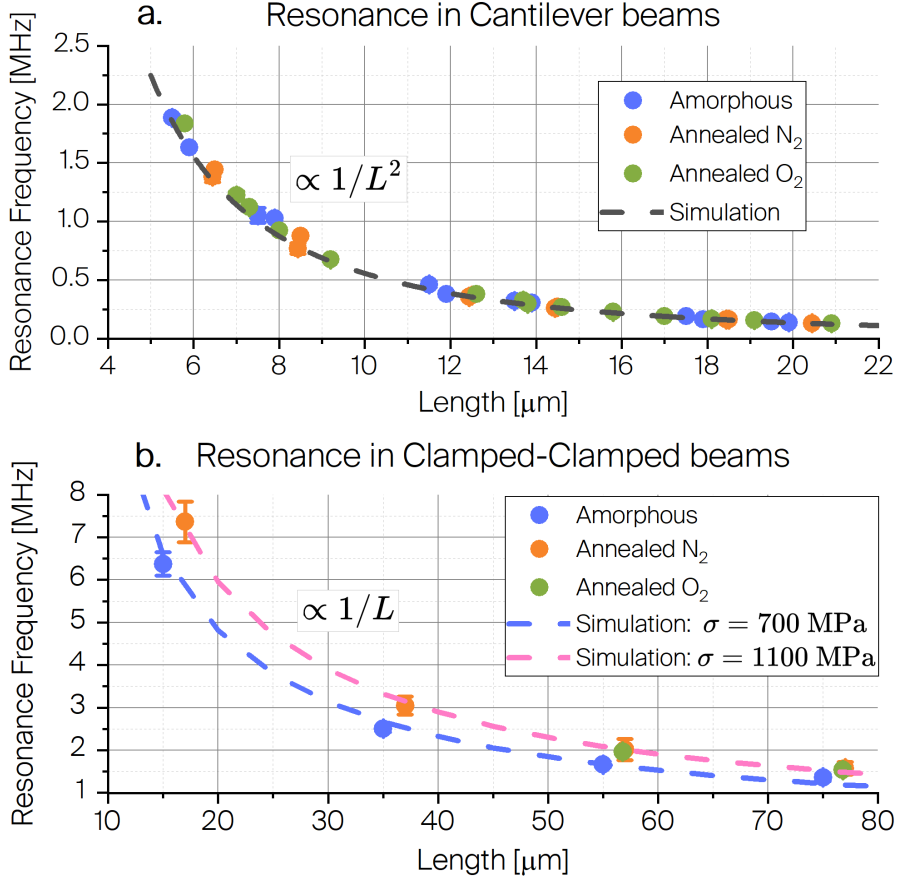

**Figure 3.** Resonance frequency analysis of cantilevers and clamped-clamped beams to investigate the effects of the crystallization process. a) Resonance frequency measurements in cantilever beams for amorphous and annealed devices (in $N_2$ and $O_2$) compared to simulations. The data for all three types of samples overlap, which means that the annealing process did not alter the (mechanical) material properties. b) Resonance measurements in clamped-clamped beams reveal a distinction between amorphous and annealed devices. The annealed devices (in $O_2$ or $N_2$) exhibit a significant increase in internal stress, approximately 60% higher than in amorphous devices, as evidenced by the comparison of the frequencies with simulations.

After proving that the annealing has changed the crystallographic phase of $HfO_2$ while preserving several other properties, we measure the flexoelectric coefficient for the three $HfO_2$ forms: amorphous, annealed in $N_2$, and annealed in $O_2$ atmosphere.

The flexoelectric coefficient is measured using a methodology described in a previously published work[13]. Importantly, this approach isolates flexoelectricity from competing phenomena such as piezoelectricity, electrostriction, and electrostatic effects. The fabricated cantilevers and clamped-clamped beams are actuated by applying a voltage signal and their displacement is read using a Laser Doppler Vibrometer (LDV). The methodology first distinguishes the combined contributions of flexoelectric and piezoelectric forces from electrostatic and electrostrictive ones by using modulated input signals. The second step involves distinguishing between flexoelectricity and piezoelectricity by analyzing the changes in resonance frequency of clamped-clamped devices. If piezoelectricity is present in the material, any applied voltage modifies the stress in the clamped-clamped beams, resulting in a significant change in resonance frequency[13,24]. If the change in resonance frequency does not occur or is of the same relative magnitude as in cantilevers, we have confirmation that the cantilever movement is purely the result of flexoelectricity.

We isolate flexoelectricity, verifying that piezoelectricity is non-existent. Additionally, the method provides values for the electrostatic and electrostrictive forces in the amorphous and annealed cantilevers. The electrostatic and electrostrictive effects remain the same across all types of devices, indicating that annealing has not influenced these two effects. Details on this can be found in the supplementary material.

Figure 4 presents the flexoelectric coefficients measured in microcantilevers of various lengths. The amorphous phase exhibits the highest coefficient, with $\mu_{eff} = 105 \pm 10 \, pC/m$, averaging the measurements of 10 cantilevers. The annealed HfO2 in a $N_2$ atmosphere has the lowest coefficient, measuring only around 25% of that of the amorphous phase ($\mu_{eff} = 26 \pm 4 \, pC/m$), with a measuring sample of 12 cantilevers. In contrast, the coefficient for the annealed HfO2 in an $O_2$ atmosphere is higher, at around 50% of the amorphous phase ($\mu_{eff} = 54 \pm 6 \, pC/m$), with a measuring sample of 7 cantilevers.

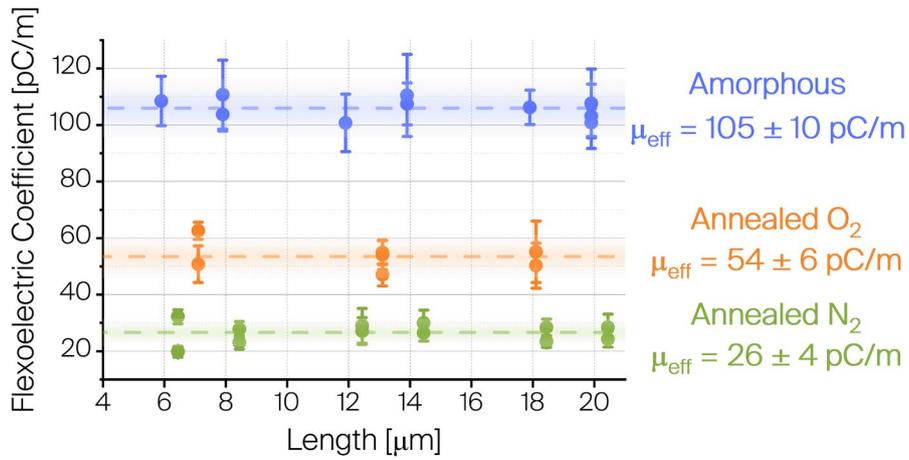

**Figure 4.** Measured flexoelectric coefficients for different devices as a function of the length of the cantilevers. The amorphous devices exhibit the highest flexoelectric coefficient, measured at 105 ± 10 pC/m. In contrast, the coefficients for annealed devices are significantly lower, with the devices annealed in an oxygen atmosphere showing a value of 54 ± 6 pC/m and the devices annealed in a nitrogen atmosphere showing a value of 26 ± 4 pC/m.

Our study shows that the annealing process reduces the flexoelectric coefficient in HfO$_2$. This result was unexpected, as our initial hypothesis was that increased structuring in the material would enhance or maintain the flexoelectric coefficient. Furthermore, the flexoelectric coefficient varies depending on the annealing atmosphere, despite providing the same crystalline form. This suggests that factors different than crystallinity might play a significant role in influencing flexoelectricity.

To explore potential explanations, we examined the relative permittivity of the samples, given that the flexoelectric coefficient depends on this property. Polarization-electric field (P-E) loop measurements confirmed non-ferroelectric behavior and similar relative permittivity across all samples ($\varepsilon_r = 24 \pm 3$), indicating that the observed differences are not due to changes in permittivity. Details on the P-E measurements can be found in the supplementary material.

The most plausible explanation, consistent with all our measurements, is the influence of oxygen vacancies in HfO$_2$. Oxygen vacancies are the most common intrinsic defect in HfO$_2$ and are currently a hot topic due to their impact on physical, electrical, and optical properties[25,26]. Oxygen vacancies are present in HfO2 even in its amorphous form, with their concentration depending on the Atomic Layer Deposition (ALD) temperature[25]. Additionally, in our case, the electrodes have a thin 2 nm Ti adhesion layer, which is known to attract oxygen atoms when in contact with HfO2[27]. Annealing the material further increases the number of oxygen vacancies[25]. Modifying the annealing atmosphere to either oxygen ($O_2$) or nitrogen ($N_2$) allows us to influence this effect—minimizing it in an $O_2$ atmosphere and increasing it in an $N_2$ atmosphere. Our experiments show that annealing in oxygen, which would produce fewer vacancies, yields a larger flexoelectric coefficient compared to annealing in nitrogen. This seems to show that oxygen vacancies degrade the flexoelectricity in $HfO_2$. The reason could lie in the crystal lattice disruptions caused by the vacancies, as well as the introduction of localized electronic states that trap charge carriers[28]. These inhomogeneous local fields can interfere with the uniform polarization response required for a strong flexoelectric effect.

In conclusion, we measure the flexoelectric coefficient for $HfO_2$ in two different phases: amorphous and tetragonal (annealed in $N_2$ and $O_2$ atmospheres). To understand the changes induced by annealing, we conduct cross-section imaging, XRD analysis, resonance frequency characterization, P-E loops, and relative permittivity measurements. These analyses reveal that annealing promotes crystallization into the tetragonal phase and increases stress within the $HfO_2$ layer. The flexoelectric coefficient measurements report the highest value for $HfO_2$ amorphous, then followed by the annealed in oxygen, and the lowest value is given by the samples annealed in nitrogen. The explanation of these results seems to lie in the oxygen vacancies generated during the annealing process which lower the flexoelectric coefficient due to lattice distortions and trapped charges. The overshadowing effect of oxygen vacancies does not allow us to distinguish the sole impact of crystallization on the flexoelectric coefficient.

**Supplementary material section**

The supplementary material details the microfabrication process for cantilevers and clamped-clamped beams made of hafnium oxide. It explains the setup for measuring the flexoelectric coefficient using a Laser Doppler Vibrometer. Annealing tests are presented to show that different temperature profiles result in the crystallization of $HfO_2$ into its tetragonal phase. Polarization-electric field (P-E) loop measurements confirm the non-ferroelectric nature of $HfO_2$ and provide data to calculate its relative permittivity.

**Acknowledgments**
The authors express their gratitude to Dragan Damjanovic and Gustau Catalan for the insightful discussions that helped explain the results. To Andrea Elisei and Andrea Bruder for the contributions on the different annealing temperature profiles.
This work was supported by financial support from the Swiss National Science Foundation via grants PP00P2_170590 and CRSII5_189967.

**Author declarations section**

The authors have no conflicts to disclose.

**Daniel Moreno-Garcia:** Conceptualization (equal); Data curation (lead); Formal Analysis (lead); Investigation (lead); Methodology (equal); Software (lead); Validation (lead); Visualization (lead); Writing – original draft (lead); Writing – review & editing (lead).
**Guillermo Villanueva:** Conceptualization (Lead); Funding acquisition (Lead); Project

administration (Lead); Resources (Lead); Supervision (Lead); Writing – review & editing (Equal).

**Data availability statement**

The data that support the findings of this study are available from the corresponding author upon reasonable request.

# SM - Effect of Annealing and Oxygen Vacancies on Flexoelectricity in Hafnia (HfO2)

*Daniel Moreno-Garcia[1], Luis Guillermo Villanueva[1]*

[1] Advanced NEMS Group, École Polytechnique Fédérale de Lausanne (EPFL), Lausanne 1015, Switzerland
E-mail: Guillermo.Villanueva@epfl.ch

## Section 1. Microfabrication of Devices

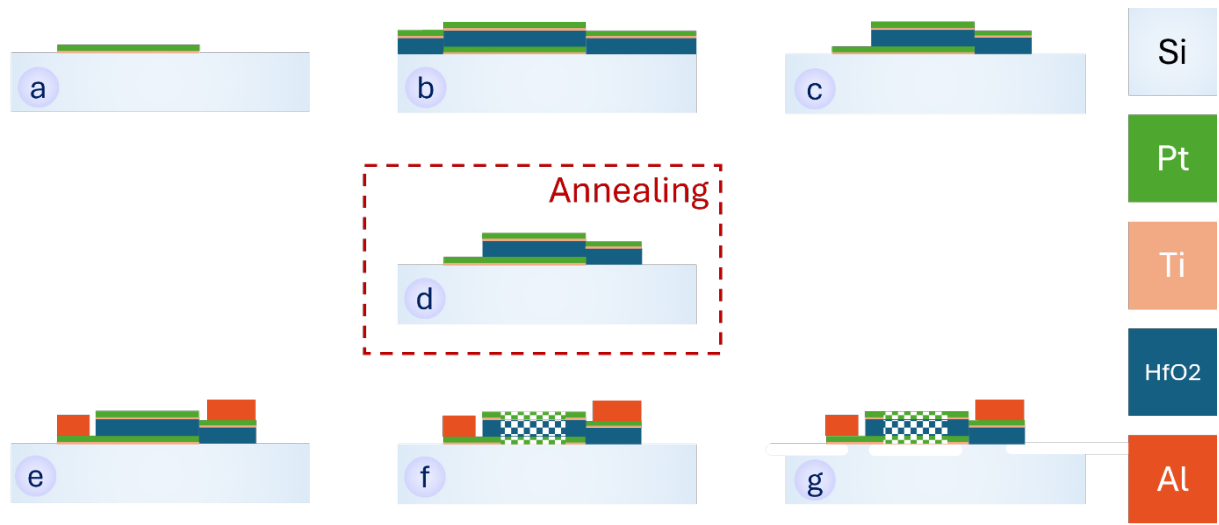

**Figure 1**. *Schematics showing the simplified fabrication steps for cantilevers and clamped-clamped beams made of hafnium oxide and platinum.* **a)** *Liftoff of an evaporated bottom electrode thin film of platinum (18 nm) with and adhesion layer of titanium (2 nm);* **b)** *ALD deposition of hafnia (50 nm) followed by top electrode evaporation (2 nm Ti + 18 nm Pt);* **c)** *Patterning and etching of the top electrode and dielectric layer;* **d)** *Annealing the samples following different temperature profiles.* **e)** *Liftoff of Aluminum pads;* **f)** *Chip-level fabrication to pattern and release the actuators through isotropic silicon etching. The checkered area indicates the locations of the cantilevers and clamped-clamped beams.*

## Section 2. Setup to measure the flexoelectric coefficient.

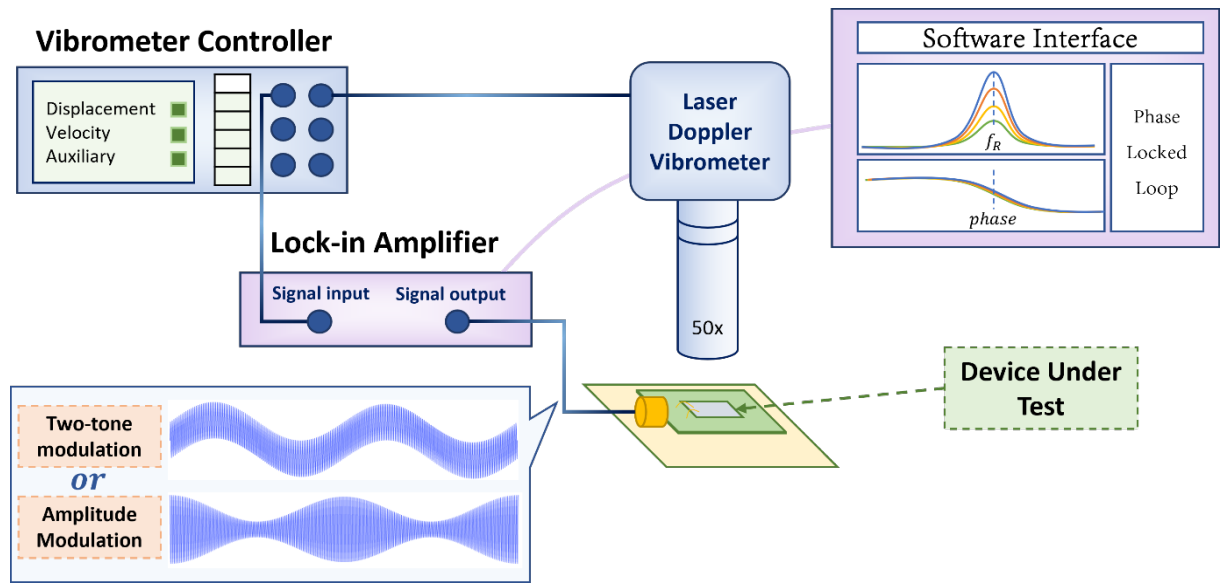

*Figure 2.* Illustration of the complete set-up for measuring the flexoelectric coefficient of cantilevers by using the converse flexoelectric effect. In this setup, a Laser Doppler Vibrometer (LDV) points at the cantilever and monitors its displacement. A lock-in amplifier actuates the cantilevers with a modulation voltage and reads the displacement data from the LDV. This data is then demodulated to obtain the effects of flexoelectricity and piezoelectricity in the beam. Additionally, the lock-in amplifier monitors the resonance frequency using a Phase Locked Loop (PLL).

## Section 3. Annealing studies

To decide the optimal annealing recipe for the samples, various tests were conducted to establish the appropriate temperature, as well as the heating and cooling rates.

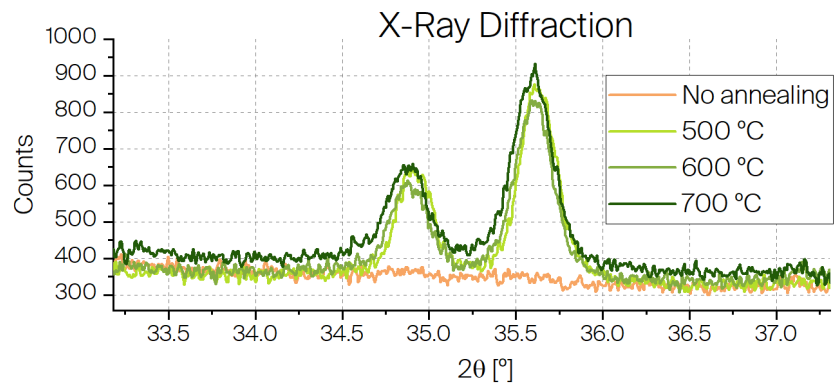

*Figure 3.* X-Ray Diffraction data for different annealing configurations, focusing on the region where crystallization peaks for hafnia are observed. This figure shows the emergence of peaks corresponding to the tetragonal phase of hafnia at all annealing temperatures (500, 600 and 700 ºC), which are absent in the amorphous phase.

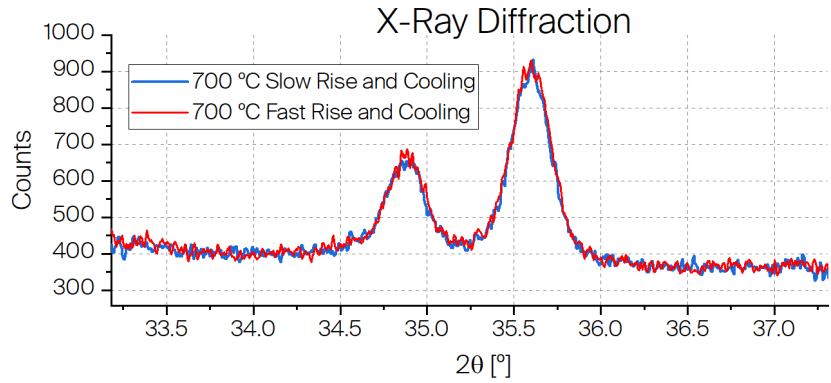

*Figure 4.* X-Ray Diffraction data for different annealing heating up and cooling down durations, focusing on the region where crystallization peaks for hafnia are observed. In one case, the sample was heated at a rate of 20ºC/min and cooled via convection for over 2 hours. In the other case, the sample was heated at a rate of 40 ºC/min and cooled using forced convection for over 1 hour. The results indicate that both methods yield identical crystallization outcomes.

## Section 4. P-E Loops

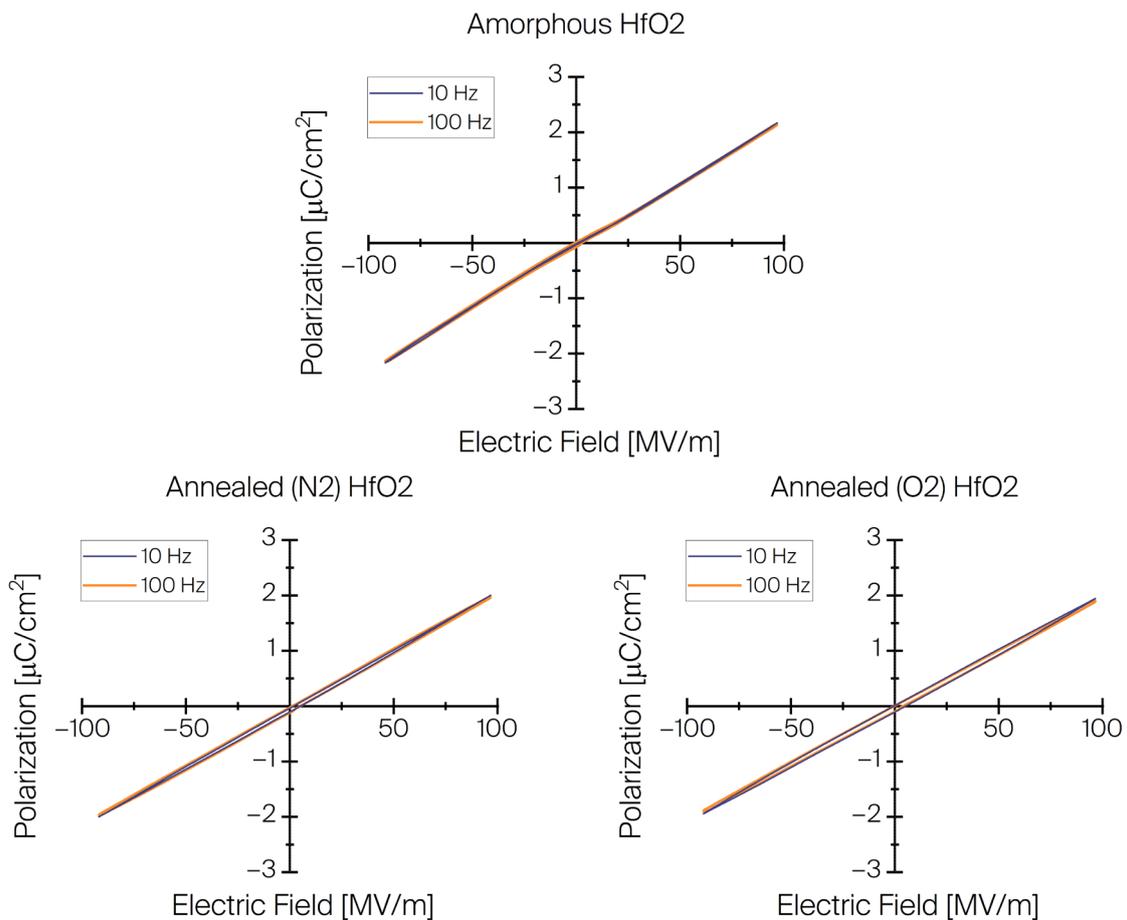

*Figure 5.* Polarization and Electric Fields (P-E) loops for the different hafnium oxide configurations: amorphous, annealed in $N_2$ atmosphere and annealed in $O_2$ atmosphere. All cases show linear loops, indicating that the material is non-ferroelectric. The only difference is a slight opening in the loops for the annealed samples, likely due to a higher leakage current from an increased concentration of oxygen vacancies.

From the P-E loop measurements, the relative permittivity of the material can be calculated using the formula:

$$\varepsilon_r = \frac{P}{\varepsilon_0 E} + 1$$

The relative permittivity values for the different cases were found to be very similar. After extensive statistical analysis, the final relative permittivity of our hafnia was determined to be:

$$\varepsilon_r(HfO_2) = 24 \pm 3$$

## Section 5. Electrostatic and electrostriction value comparison

The contributions of electrostatic and electrostriction effects are separated from those of flexoelectricity and piezoelectricity. The electrostatic and electrostrictive effects are very similar in both amorphous and annealed samples, as illustrated in Figure 6.

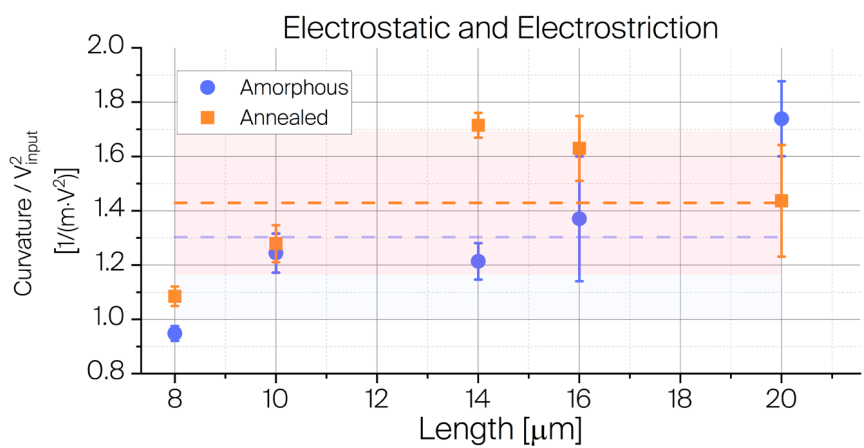

*Figure 6. Curvature normalized by the cantilever's input voltage, induced by electrostatic and electrostrictive effects in cantilevers of varying lengths. The measurements show a very similar value for amorphous and annealed samples, indicating that changes in crystallography do not significantly impact the electrostatic and electrostrictive effects.*

From Figure 6, the average curvature induced by electrostatic and electrostrictive effects is determined to be:

| Amorphous: $1.4 \pm 0.3 \ [1/m \cdot V^2]$ | Annealed: $1.3 \pm 0.3 \ [1/m \cdot V^2]$ |
|---|---|